%% file: 0_main.tex
\RequirePackage{rotating}

 \documentclass[sigconf,anonymous=false]{acmart}

\AtBeginDocument{%
  \providecommand\BibTeX{{%
    \normalfont B\kern-0.5em{\scshape i\kern-0.25em b}\kern-0.8em\TeX}}}

\copyrightyear{2024}
\acmYear{2024}
\setcopyright{rightsretained}
\acmConference[CUI '24]{ACM Conversational User Interfaces 2024}{July 8--10, 2024}{Luxembourg, Luxembourg}
\acmBooktitle{ACM Conversational User Interfaces 2024 (CUI '24), July 8--10, 2024, Luxembourg, Luxembourg}\acmDOI{10.1145/3640794.3665580} \acmISBN{979-8-4007-0511-3/24/07}

\usepackage[utf8]{inputenc}
\usepackage{blindtext}
\usepackage{soul}
\usepackage{rotating}
\usepackage{multirow}
\usepackage{multicol}
\usepackage{array}
\usepackage{booktabs}
\usepackage{subcaption}

\title[(Social) Trouble on the Road]{(Social) Trouble on the Road: Understanding and Addressing \\Social Discomfort in Shared Car Trips}
\author{Alexandra Bremers}
\affiliation{%
  \institution{Cornell Tech}
  \city{New York, NY}
  \country{USA}}
\email{awb227@cornell.edu}

\author{Natalie Friedman}
\affiliation{%
  \institution{Cornell Tech}
  \city{New York, NY}
  \country{USA}
}

\author{Sam Lee}
\affiliation{%
  \institution{Cornell Tech}
  \city{New York, NY}
  \country{USA}
}

\author{Tong Wu}
\affiliation{%
  \institution{Rutgers University}
  \city{New Jersey, NJ}
  \country{USA}
}

\author{Eric Laurier}
\affiliation{%
  \institution{University of Edinburgh}
  \city{Edinburgh}
  \country{United Kingdom}
}

\author{Malte Jung}
\affiliation{%
  \institution{Cornell University}
  \city{Ithaca, NY}
  \country{USA}
}

\author{Jorge Ortiz}
\affiliation{%
  \institution{Rutgers University}
  \city{New Jersey, NJ}
  \country{USA}
}

\author{Wendy Ju}
\affiliation{%
  \institution{Cornell Tech}
  \city{New York, NY}
  \country{USA}
}

\renewcommand{\shortauthors}{Bremers, et al.}

\begin{abstract}
Unpleasant social interactions on the road can negatively affect driving safety. At the same time, researchers have attempted to address social discomfort by exploring Conversational User Interfaces (CUIs) as social mediators. Before knowing whether CUIs could reduce social discomfort in a car, it is necessary to understand the nature of social discomfort in shared rides. To this end, we recorded nine families going on drives and performed interaction analysis on this data. We define three strategies to address social discomfort: contextual mediation, social mediation, and social support. We discuss considerations for engineering and design, and explore the limitations of current large language models in addressing social discomfort on the road. 
\end{abstract}



\begin{document}

\maketitle

\input{1_introduction}
\input{2_relatedwork}

\input{3_methods}

\input{4_results}
\input{5_discussion}
\input{6_limitationsandfuturework}
\input{7_conclusion}
\input{8_acknowledgements}

\bibliographystyle{ACM-Reference-Format} 
\bibliography{references}

\input{9_appendix}

\end{document}

%% file: 1_introduction.tex
\section{Introduction}

Cars aren't designed with sociability in mind. Unpleasant social interactions in the car bear negative consequences whose impacts go far beyond the enjoyment of the drive. In fact, concerns about expected social awkwardness are one factor disincentivizing people from taking more ecologically efficient shared rides \cite{NIELSEN2015113}. Furthermore, in an analysis of 856 serious casualty car crashes, \citet{BEANLAND201399} found that driver inattention was a major aspect of most crashes, and was commonly caused by fatigue, intoxication, or distraction, including distractions from passengers. 

Could cars be designed to reduce social discomfort? Historically, automotive interface research focused on physical causes of discomfort (such as uncomfortable seats \cite{zhang1996identifying, akamatsu2013automotive, shen1997redefining}, motion sickness \cite{kuiper2020knowing, dong2010postural, wada2016motion}, or characteristics related to AC, noise, space, or smell \cite{constantin2014elements}), or psychological discomfort (e.g. driver stress \cite{paredes2018fast}). At the same time, interaction research in vehicles emphasizes examining driver and passenger activities in context, in the field, and under as natural a scenario as practicable \cite{meschtscherjakov2011capture}. This can involve instrumenting the vehicle with video cameras to record journeys (e.g., \cite{mackay1988video,semmens2019now,laurier2008driving,baltodano2015rrads}) and then perform detailed ethnographic, conversational, or interaction analysis to understand what is occurring \cite{SUCHMAN1995233,jordan1995interaction, McDonald}. A growing body of work uses ethnomethodological approaches to describe the experience of distinct groups traveling together in a car (e.g., \cite{laurier2008driving,laurier2004doing,brown2012normal}).

As Conversational User Interfaces (CUIs) are becoming increasingly present in cars (studied in e.g., \cite{large2019smalltalk,jestin2022effects}), this presents both challenges and opportunities regarding their effects on social discomfort. On the one hand, interruption timing will be important to ensure smooth interaction \cite{semmens2019now,caber2023driver}. On the other hand, CUIs could potentially, just like prior work on robots has shown, be explored for intervening in conflicts \cite{jung2015moderate} and improving inclusivity \cite{tennent2019micbot} to achieve higher social comfort. Furthermore, developments in Large-Language Models (LLMs) add to the possibilities and impacts of CUI interactions (see \citet{skjuve2023chatgpt}, and ethical considerations described by \citet{bang2023examination}).

Our contributions in this paper are: 1) a user study and construct of on-road social discomfort, 2) the conceptualization of three mitigation strategies, and 3) a reflection on what this implies for CUIs against a backdrop of increasing capabilities of LLMs.

%% file: 2_relatedwork.tex
\section{Related Work}
\subsection{Defining discomfort in cars}
Automotive human factors research typically focuses on \textit{physical} sources of discomfort, such as uncomfortable seats \cite{zhang1996identifying, akamatsu2013automotive, shen1997redefining}, motion sickness \cite{kuiper2020knowing, dong2010postural, wada2016motion}, or aspects of AC, noise, space, or smell \cite{constantin2014elements}. In ergonomic studies, discomfort is related to the position of and pressures on the human body \cite{dul1994ergonomic}. In cognitive science research, discomfort comes from feelings of cognitive dissonance \cite{elliot1994motivational}, while in social psychology research, discomfort can come from psychological distance from other people \cite{zhang2019interaction}. 

Much of the literature about social discomfort discusses social anxiety \cite{leary1997social}, social exclusion \cite{macdonald2005does} or rejection \cite{eisenberger2003does}, which could happen in cars but is not necessarily specific to the enclosed space of the vehicle. However, the vehicle can definitely be seen as a social space, as exemplified and described in ethnomethodological works (\cite{laurier2008driving,laurier2004doing,brown2012normal}). We aim to combine these backgrounds and investigate sociability and social discomfort, specifically in the space of the car.

\subsection{Capturing discomfort in cars}
First, discomfort can be elicited by introducing uncomfortable stimuli, such as heavy metal music and math problems \cite{paredes2018fast}, which are then correlated with sensor data. Another approach is to collect naturalistic data and filter this data afterward for moments of interest based on sensor changes (e.g., \citet{seacrist2018analysis}). Various types of sensors and algorithms can detect driver stress \cite{gulian1990stress, paredes2018fast}, or anticipate anxiety people might feel, for example, around electric vehicle range \cite{rauh2015understanding, franke2012experiencing,jung2015displayed}. 

Second, experiential surveys can be used to account for the subjective perspective. Some standardized scales for evaluating discomfort include the motion sickness awareness questionnaire (MSAQ)~\cite{gianaros2001questionnaire} or NASA TLX workload questionnaire \cite{hart1988development}. One downside of these surveys is that they suffer from being administered post-hoc. To overcome this, \citet{semmens2019now} have experimented with using voice agents in situ to ask people in real-time to label moments as good or bad moments to be spoken to. 

Third, qualitative studies capture a richer aspect of the social dynamics of what occurs in cars. \citet{forlizzi2010should} used interaction videos and interviews to study collaboration during navigation. \citet{10.1145/1357054.1357316} used observations and interviews to study how GPS navigation impacts our ability to engage with our surroundings. \citet{laurier2012means} looked at how navigation occurs in family vehicles. \citet{meschtscherjakov2011capture} used contextual inquiry, ethnographic study, and cultural probes to capture the holistic experience of being in cars. \citet{laurier2008driving} goes the furthest to focus on the social dynamics within the car, to see how the automotive context alters the organization of family, colleague, or friend relations. 

\subsection{Addressing discomfort in cars}
Past efforts to reduce car discomfort often focused on seat ergonomics, the in-cabin space, thermal factors, and the acoustic environment \cite{pywell1993automotive,mohamed2007automotive, guan2003investigation, da2012combining, constantin2014elements}. Work from \citet{hassenzahl2017s} has argued for bringing an interaction design perspective focused on \textit{well-being} to the automotive space, although this work emphasizes primarily human-technology interactions, and less so the interactions between people. Works in the field of human-robot interaction (e.g., \citet{tahir2020user,noguchi2020personality,gillet2020social}), as well as conversational interfaces (e.g., \citet{kowatsch2021conversational}), have looked at how technologies could mediate during conflicts and in group interactions. This signals the potential for (conversational interface) technology to intervene in in-car interactions.

%% file: 3_methods.tex
\section{Approach}

\input{3b_userstudy}

\input{3d_dataanalysis}

%% file: 3b_userstudy.tex
\subsection{Data collection}

We recruited families through social media, notice boards, and e-mail. \footnote{The rationale for recruiting families and households was to have social interaction between people while minimizing the introduction of new social contact for COVID-19 social distancing reasons.}\footnote{Recruitment proved challenging, with the most common reasons for decline being the pandemic (fear of infection with COVID-19, having left the city, no purpose of a driving trip due to closures) or personal reasons (not having a driver's license or car, unwillingness to rent a car, distance from car rental company, not living with family).} We instrumented the car with cameras facing all occupants and the road ahead. One phone was placed under the center console for audio recording. Participants provided consent (or child assent) and agreed to have their images published. Participants then went freely on a drive of 1-2 hours. Each family was provided with a US\$50 gift card and reimbursement for rental vehicle costs (if applicable). The research was approved by \anon{Cornell IRB \#1909009034.}

\label{sec:missing}
We collected 23 hours of data from 9 families in the New York City area in spring 2021 (see \autoref{tab:dataset}). As the vehicle rides were long and featured stopping, instrumentation sometimes failed (e.g., drained batteries) or was altered by participants. \footnote{We could not have a researcher ride along in the vehicles due to concerns over COVID-19 transmission.} Nevertheless, the dataset shows a wide range of in-vehicle interactions. Synchronized videos were transcribed and translated to English where applicable. We used a spreadsheet and an online whiteboard to discuss excerpts, and created graphic transcripts (cf. \citet{laurier2014graphic}).

%% file: 3d_dataanalysis.tex
\subsection{Data analysis}
Using interaction analysis, as specified by \citet{jordan1995interaction}, groups of researchers extract richer insights from the observed interactions. Usually, video analytic approaches focus on detailed analysis of action sequences, for instance, as seen in \cite{laurier2008driving, laurier2012means}. This video capture and careful analysis can help researchers develop an intricate and grounded understanding of how these interactions unfold, providing rich insights for system design \cite{mackay1988video}. 

We applied video interaction analysis (see \cite{SUCHMAN1995233}) during weekly hour-long meetings over eleven months, with asynchronous work between meetings. One researcher first watched three participants' videos and selected excerpts that caught their attention. Through group discussions, the construct of discomfort in family car trips was set as \textit{"a situation that is negative, undesired, or captures attention in a negative way"} -- the opposite of a neutral or pleasant state. From this, we observed different kinds of discomfort, of which some were physical or ergonomic (for instance, needing to get into strange physical positions to retrieve earbuds that were dropped between the car seat). Some were psychological (feeling annoyed with a child who keeps asking when the destination will be reached) or cognitive nature (being overloaded by different types of stimuli from the car and environment). 
\footnote{We did not attempt to strictly delineate between different categories of discomfort in absolute terms but rather provided a tool for labeling and discussion.}
Then, all videos were re-analyzed to label similar moments accordingly. \footnote{The entire dataset was divided among the research team, with each researcher watching at least two participants' data.}

Of particular interest was the observation of social discomfort -- an unpleasant experience relating to interacting with others. This includes embarrassment, conflict, disagreement, ignoring, mean behavior, breaking social norms, and unpleasant social acts. We identified 116 moments of social discomfort (see \autoref{tab:analysiscounts}).

%% file: 4_results.tex
\begin{figure*}
  \centering
  \includegraphics[width=0.95\textwidth]{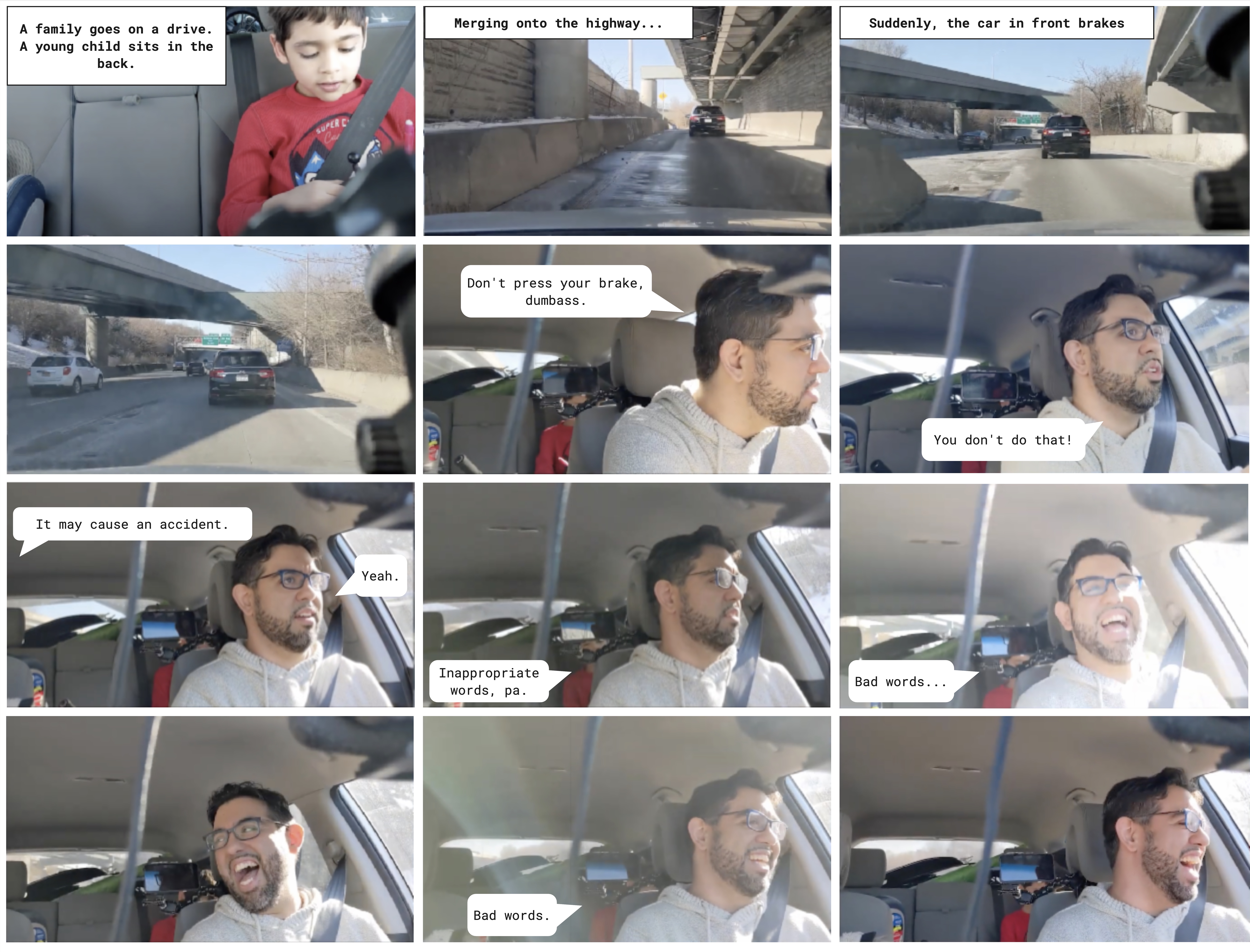}
  \caption{\textbf{Excerpt 1. }Visual transcript of a driver reacting to a car suddenly braking in front.}
  \label{fig:inappropriate}
  \Description{A visual transcript with speech bubbles of an interaction. A family goes on a drive. A young child sits in the back. Merging onto the highway… Suddenly, the car in front brakes. The dad ends up calling out the driver in front for being a dumbass, but gets corrected by his child for using "bad words". Everyone laughs.}
\end{figure*}

\section{Findings}

\subsection{High-level observations}
\label{sec:highlevelres}

\begin{figure*}
  \includegraphics[width=0.95\textwidth]{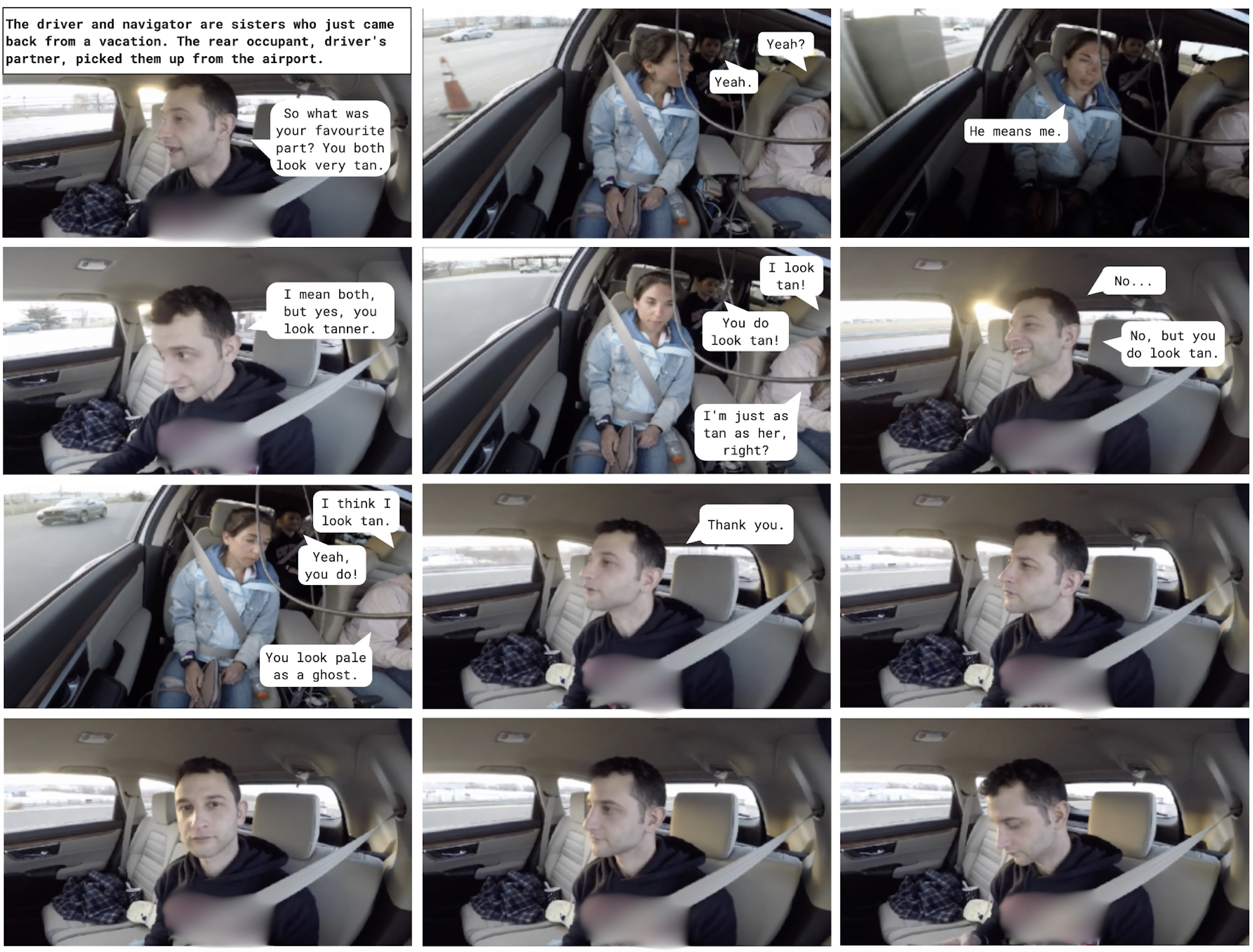}
  \caption{\textbf{Excerpt 2. }Visual transcript of a compliment resulting in a squabble.}
  \label{fig:pale}
  \Description{Visual transcript of an interaction with speech bubbles. The driver and navigator are sisters who just came back from a vacation. The rear occupant is the driver's partner, who picked them up from the airport. There is an argument about who is the tannest, and the driver calls the rear passenger "pale as a ghost".}
\end{figure*}

Interactions sometimes occurred between the occupants of a particular row, other times between all members of the vehicle, and sometimes with external people, such as during phone calls, reacting to other road users, or even communicating to the research team by directly seeking interaction with the recording equipment. Interactions were casual: rear passengers would sometimes fall asleep; and sometimes people would talk or sing to themselves. Even though the space was shared, participants sometimes closed themselves off from others by wearing headphones. Despite the constraints of the vehicle's interior and seat belt, we observed many changes in body pose, such as a navigator reaching out to hand something to a passenger in the back. We also found interactions like hugs and people leaning on each other. We selected three key moments of social discomfort (see \autoref{fig:inappropriate}, \autoref{fig:pale} and \autoref{fig:interruptions}), representing different causes and mitigation strategies, which we will discuss further in the next sections. 

\subsection{"Don't press your brake, dumbass!"}
\label{sec:excerpt1}

Excerpt 1 (\autoref{fig:inappropriate}) shows a driver/father and navigator/mother. A young child and an infant are in the back. The child is entertaining himself with a drawing demonstration. At one point, the driver merges onto the highway and is surprised by a vehicle braking before him. The driver berates the other vehicle, saying, "Don't press your brake, dumbass!" His exclamation is initially met by a supportive statement from the navigator. Then, the child chides his father for using "inappropriate words," which results in laughter and lightening of the mood in the car. While these directives from the driver ("Dumbass!") are in response to the driver in front, of course, the driver in the vehicle in front of the participant is not expected to be able to hear him. In using directives, the driver, as a father, is looking for empathy from the other car occupants as his family members -- most likely the navigator. Meanwhile, their child in the back finds a markedly different hearing of his father's words, not as a driver justifiably exasperated with poor driving from another road user but as a family member. His switch of perspective provides a category disjuncture that both parents find funny.

\subsection{"You look pale as a ghost"}
In Excerpt 2 (\autoref{fig:pale}), the driver and navigator are sisters on a long journey back from a vacation; the passenger (driver's partner) picked them up and has transitioned to sitting in the back seat. There was potential for tension in the vehicle at the outset; the person performing pick-up was late getting to the airport, and the travelers were tired. The driver has difficulty navigating and encounters numerous toll booths. %. 
A conversation ensues regarding annoyance about the drive. At this moment, the rear passenger (the first frame of \autoref{fig:pale}) asks the other occupants what their favorite part of the vacation was, adding that they both look very tanned. What follows in Excerpt 2 is a squabble ending with the rear occupant expressing discomfort through his facial expressions and even directly looking into the camera, likely conscious of being video recorded. 

The preceding context from the earlier experiences of the car occupants provides the backdrop to the emergence of discomfort during the excerpt. Instead of the invitation to storytelling and the paying of a compliment being taken at face value, the earlier experiences lead to discomfort in the moment. A compliment made by the partner as a passenger to two distinct potential recipients ("You both look very tan!") is a classic example of the fragility of paying compliments in a group situation. To compliment one person does things to the other people who are present. The compliment initially goes well until the navigator uses the compliment as a way to tease her sister, who is driving, which turns the mood around. Interestingly, after commenting, "he means me,"  the navigator stays out of the rest of the exchange, where the driver insults the passenger for looking "pale as a ghost." The passenger driving is drawn into the sibling rivalry; where we might expect him to say his partner looks more tanned as part of favoring her, he instead favors her sister and tries to mitigate that by adding, "But you do look tan". The discomforting moment reaches its peak when the partner (driver) then formulates the passenger's skin color: "You look pale as a ghost." While it is a factual description of his comparison of her tan as less than her sister's, in its response it shows something is amiss in their encounter in the car. She has been hurt in a way yet to be said, and he, having begun with a compliment, finds himself at a loss for words.

\begin{figure*}
  \includegraphics [width=0.8\textwidth] {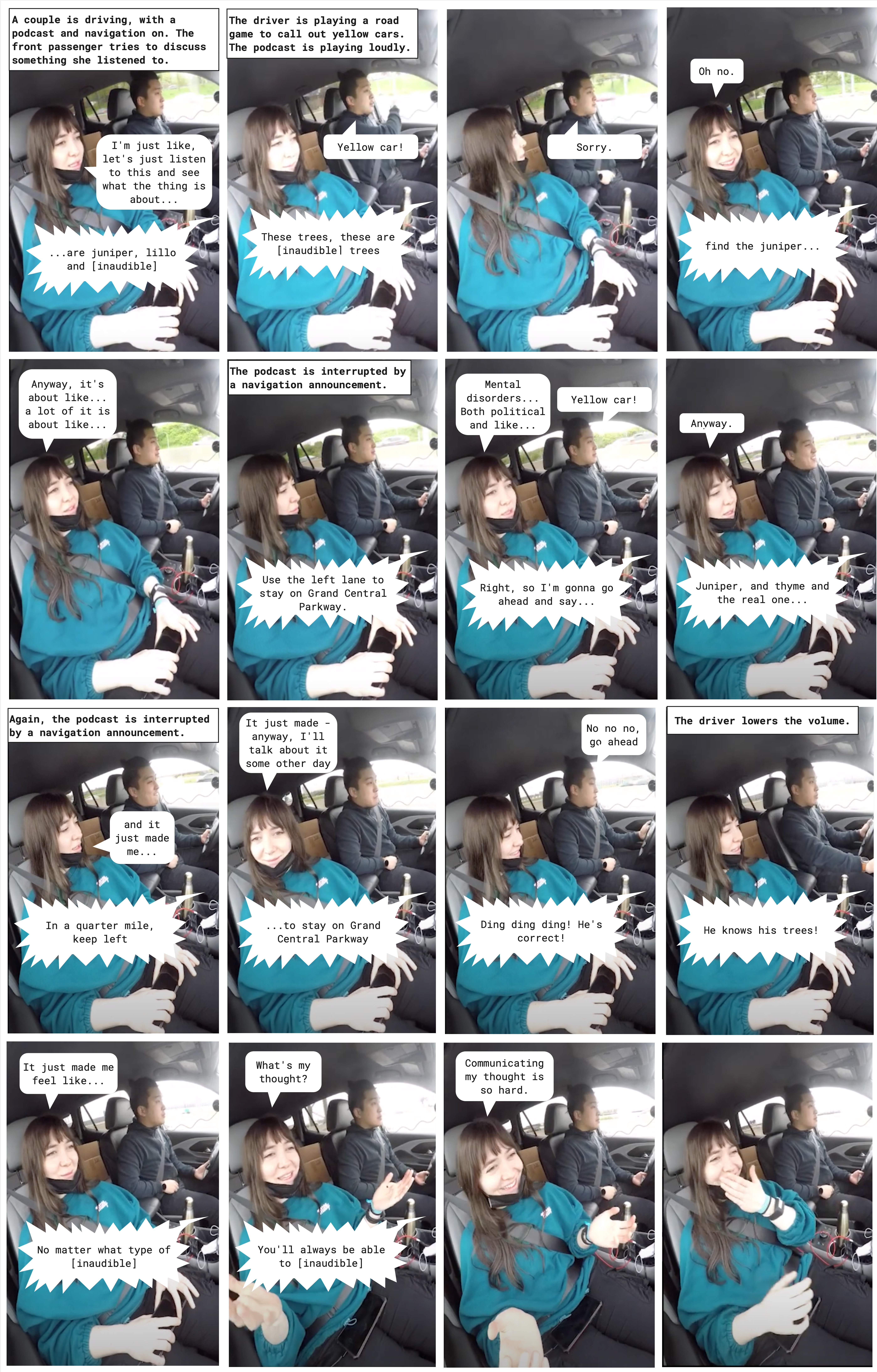}
  \caption{\textbf{Excerpt 3. }Visual transcript of a person getting repeatedly interrupted.}
  \label{fig:interruptions}
  \Description{A visual transcript with speech bubbles of an interaction. A couple is driving with a podcast and navigation on. The front passenger tries to discuss something she listened to. She keeps getting interrupted by the podcast, the navigation system, and the driver who calls out yellow cars until she loses her train of thought.}
\end{figure*}

\subsection{"Communicating my thought is so hard."}

In Excerpt 3 (\autoref{fig:interruptions}), a couple drives to run errands. Throughout the drive, a podcast plays via Bluetooth from a phone connected to the car. The phone also runs a navigation application, occasionally interrupting the podcast to announce new directions. In this excerpt, the navigator attempts to discuss an audiobook she had been listening to before the drive. While she is speaking, she is interrupted multiple times by the driver playing a road game (calling out yellow cars as he sees them), the podcast (a conversation between two people), and the car navigation application, which is also interrupting the podcast. The volume of the podcast and navigation is about equal to that of the navigator's voice. At one point, she says that she will continue to talk about it another day, upon which the driver encourages her to resume her discussion. The podcast loudly interrupts, and the driver lowers the volume. However, the navigator now expresses that she has lost her train of thought. Multiple activities and interruptions are at play in this situation. Interestingly, in this case, the demands on the passenger seem higher than for the driver. The frequent interruptions led to the passenger giving up her discussion in the face of the driver's continued playing of the game and the noise from the podcast and the navigation app. After the initial interruption, when the driver calls out "yellow car" as a move in the game he's playing, his partner looks at him and then looks away and attempts to restart her talk with, "Anyway...".  In doing so, she restarts while showing further signs of the effort required to formulate her thoughts about the story she has listened to. Her next interruption is from the GPS, which she stops talking through, given its driving relevance. When she attempts to restart for a third time, talking over the talking book, she is interrupted by a second announcement from the GPS. The driver only lowers the podcast volume after his partner abandons trying to discuss her podcast. Interestingly, neither the driver nor the navigator turned the volume down before this point. 

%% file: 5_discussion.tex
\section{Discussion}

\subsection{How did participants address social discomfort?}

\subsubsection{Successfully diverting attention}
The interaction in Excerpt 1 shows us a driver who is initially upset by a violation of social norms by a rude road user -- when encountering incivility, it is customary to be surprised and look for empathy in others \cite{smith2010incivility}. When the child makes a funny statement that leads to laughter, it might be interpreted as an example of situational change and attentional deployment from the model of emotion regulation by \citet{gross1999emotion}. The mood in the car changes back to positive after the child calls out the dad for using inappropriate language. Apart from the funny remark of the child, the fact that the navigator agrees with the driver that the other road user could have caused an accident is an example of a successful bid for connection in a stressful scenario by expressing understanding \cite{atkinson2005emotional}. 

\subsubsection{Unsuccessfully diverting attention}
In other cases, intervening to mitigate discomfort only worsens matters. This class of observations tended to be less often present in the dataset, yet sheds light on situations that are tricky to navigate, for both people and machines. Excerpt 2 exemplifies what happens when mitigation fails. The compliment represents a bid for connection as described by \citet{gottman2001relationship}, where the partner's response, in this case, is to turn away. According to Gottman's studies, the likelihood of someone making another bid after a rejected bid will be low. In a situation like this, one may ask oneself if it is better to intervene, or wiser to stay quiet and at least do no harm. 

\subsubsection{Regulating the space of interaction}
There were also cases where discomfort seemed to originate from a complex combination of sources. For example, the car itself often caused distractions that would add to the number of interactions already going on inside. In the example of Excerpt 3, the navigator attempts to connect to the driver and fails due to an overload of other signals from podcasts and navigation that interrupt her. When the driver finally turns the volume of the podcast down, this can be interpreted as an acceptance of the bid of the navigator. 

\subsection{Strategies for addressing social discomfort}
\label{sec:strategies}
\subsubsection{Strategy A: Contextual Mediation}\hfill\\
The general malaise in Excerpt 2 probably had a lot to do with the physical tiredness of the occupants after a long journey. It is often difficult to pinpoint the "cause" of socially uncomfortable moments because one type of discomfort often becomes another. Recognizing this, it becomes clearer how seemingly superfluous comfort features, which help to manage the in-cabin environment--the climate, sound level, or distraction--could potentially have ramifications for the safety and social well-being of all the people in the vehicle by contributing to the overall experience. Luckily, contextual mediation is widely present in present-day vehicles, apparent through many features aiming to improve physical comfort: from massage seats to individualized climate control (not just for air circulation but also localized in the steering wheel or backrest), mini-fridges, entertainment displays, and directional audio. Contextual mediation features do not need to relate to the situation of discomfort directly and can improve a situation without requiring complex and sensitive responsiveness to interactions or risking making matters worse (such as in \autoref{fig:pale}). We thus recommend contextual mediation as the initial strategy to mitigate social discomfort -- this solution could take place before using CUIs.

\subsubsection{Strategy B: Social Mediation}\hfill\\
In Excerpt 3, it is clear that too many things are going on simultaneously in the car--the audio environment is so taxing that the passenger completely loses her train of thought--and the driver tries to intervene but also adds to the problems himself. To aid in people's emotional regulation, the car could proactively mute the podcast, time the navigation messages or use peripheral displays to signal upcoming directions or provide socio-metric cues to the driver that he is being unhelpful \cite{seppelt2018assessing, 10.1145/3546726, de2004exploring, choudhury2003sensing}. This social mediation approach could work in scenarios where the cause of discomfort is relatively straightforward, such as the volume level and interruptions in Excerpt 3. 

\subsubsection{Strategy C: Social Support}\hfill\\
Lastly, the strategy of social support is the most complex, risky, and sophisticated way to mitigate social discomfort. We picked two excerpts that show how this can go well and how this can go badly. Excerpt 1 shows an example of success in intervention through social support, as the navigator and backseat passenger help change the driver's mood from hostility to happiness by offering humor and empathy in response to the driver's expressed feelings. Conversely, Excerpt 2 shows how attempting social support can go completely awry and fail to lead to a positive mood in the vehicle. This excerpt also highlights how complicated it can be to provide social support and that there are limitations to what machines can do to help -- as we see, even people sometimes lack the finesse needed to provide successful social support.

\subsection{Next steps for engineering and design}

\subsubsection{How would ChatGPT intervene?}\hfill\\
Of our three mediation strategies presented in \autoref{sec:strategies}, social mediation and social support would specifically involve opportunities for conversational user interfaces. To investigate the status quo of LLMs' capabilities to address discomfort, we asked ChatGPT 4.0 \cite{OpenAI_ChatGPT_4} how it would intervene if it were a voice assistant tasked with reducing social discomfort in a car (partially shown in \autoref{tab:gpt}, full prompt and response in Appendix). Looking at these responses -- entertaining, yet also mildly condescending -- it is clear that ChatGPT still needs to work on its social finesse before it can effectively perform the social support strategy. Some of this social finesse could come from contextual information providing an increased awareness of what is going on in the car, which, fortunately, many technical methods can already help with.

\begin{table*}[ht]
\centering
\caption{ChatGPT 4.0 acting as a voice assistant that reduces social discomfort (full details in Appendix)}
\begin{tabular}{>{\raggedright\arraybackslash}p{0.23\linewidth} >{\raggedright\arraybackslash}p{0.72\linewidth}}
\toprule
\textbf{Uncomfortable interaction} & \textbf{How ChatGPT would intervene} \\
\midrule
"Don't press your brake, dumbass!" & "It sounds like we had a bit of a surprise on the road just now. It's always challenging when unexpected things happen while driving, but it's great how everyone stays alert. Remember, it's important to keep the atmosphere positive for everyone in the car, especially our younger passengers. How about we play a quick game to lighten the mood? Let's see who can spot the most interesting thing outside!" \\
\addlinespace
"You look pale as a ghost" & "Speaking of vacations, it sounds like you all had a fantastic time soaking up the sun! I've heard many people bring back not just tans, but also interesting stories and souvenirs from their travels. Do you have any memorable experiences or special items you brought back from your trip that you'd love to share?" \\
\addlinespace
"Communicating my thought is so hard" & "If we're enjoying the game of spotting yellow cars and also want to discuss the podcast, maybe we can do both. Let's designate a specific time for each activity. Does that sound good?" \\
\bottomrule
\end{tabular}
\label{tab:gpt}
\end{table*}

\subsubsection{Multi-modal contextually aware CUIs}\hfill\\

If our eventual goal is to have in-vehicle conversational agents contribute to mediating social discomfort, we believe a deeper understanding of how social discomfort could be detected and modeled is essential. Relying on conversation alone does not give enough information to be able to determine the right response. Various cues and data sources can give away social discomfort. For example, discomfort might be identified by the front passenger's body position and facial features in Excerpt 3. The cause of the discomfort, however, is the frequent interruptions that disrupt the cadence of the conversation. To detect that phenomenon, a model of the passenger's speech patterns might be needed to detect moments when she changes her talking.

Our video corpus highlighted numerous opportunities for in-vehicle agent interaction; there are extended periods where little is exchanged, and vehicle participants seem bored and appear to welcome opportunities for interaction. However, it is also easy to see that poorly timed interactions can harm the in-vehicle social dynamics. Earlier research indicates that car phone conversations are safer when the remote conversationalist has the context of the traffic the driver is driving through \cite{schneider2005calling}. Similarly, intelligent communication timing, where conversation agents are combined with models of the real-time context of the car, could be the first step toward socially viable in-vehicle conversation. Prior work looks at when drivers are interruptible \cite{semmens2019now, wu2021learningto, caber2023driver}, or studies the effect of in-vehicle agent interactions \cite{jeon2015effects} under relatively controlled settings. 

Finally, personal and cultural differences should be taken into account in the design of social discomfort interventions. Prior research on CUIs in cars has found that personalization regarding culture and context could improve the interaction with in-car voice assistants \cite{braun2019your}.

\subsection{Will technology solve social discomfort?}\label{sec:techrole}
While social discomfort is often unpleasant, conflict is necessary for maintaining relationships and personal growth. Yet, the car is not always the best place to resolve conflict -- being on the road, safety is the priority, and uncomfortable situations could threaten attentiveness to driving. There is thus a tension between, on the one hand, the view to leave human relationships alone -- especially close and intimate ones -- yet, on the other hand, the reality is that conflicts on the road could threaten driving safety (see \citet{BEANLAND201399}). 

We advise that designers consider their role in these processes before choosing their stance and the strategies they employ to address social discomfort, as artificial intelligence technologies can have real effects on human interactions and relationships \cite{hohenstein2020moral}. Nonetheless, we advise the automotive research community to consider the role of cars in their wider cultural and emotional context, which includes the role of in-car interactions in shaping human interactions (see \citet{sheller2004automotive}). 

%% file: 6_limitationsandfuturework.tex
\section{Limitations}

Despite its common use in interaction analysis, video fundamentally is a transformation of reality that can never capture \textit{all} its fullness \cite{jordan1995interaction}. Second, although it is common for interaction analysis studies to feature a similarly sized participant pool from a limited area (see, e.g., \cite{eilitta2021children,laurier2008driving}), we would expect to find different nuances if the same study were run in another location. 

%% file: 7_conclusion.tex
\section{Conclusion}

We collected and analyzed a video corpus to understand discomfort in shared rides. Through three selected excerpts, we discuss the nature of social discomfort in vehicles. We present opportunities for intervention as three strategies consisting of contextual mediation, social mediation and social support. Finally, we discuss fundamental limitations and considerations that should be kept in mind when attempting to address the problem of discomfort in cars through technology -- especially generative AI-based conversational interfaces such as those incorporating ChatGPT. 

%% file: 8_acknowledgements.tex
\begin{acks}
    The authors would like to thank members of the \anon{Cornell Tech} research community for supporting this work. The work was made possible in part by research funding from \anon{Nissan Motor Co. Ltd}.
\end{acks}

%% file: 9_appendix.tex
\appendix

\section{Appendix}
\begin{table*}[h]
\small
\caption {Overview of the Family Car Trip dataset} \label{tab:dataset} 
\begin{tabular}{ |p{1.9cm}||p{2.4cm}|p{2.4cm}|p{2.4cm}| p{2.4cm}|  }
 \hline
 \multicolumn{5}{|c|}{Characteristics per participant} \\
 \hline
Participant& Journey & \# People & Language & Duration\\
 \hline
P1 & House visit & 4 & Spanish & 1:40:45 \\
P2 & Ski Trip & 7 & Hebrew & 2:25:47 \\
P3 & Errands & 2 & English & 2:56:05 \\
P4 & Airport pickup & 3 & English & 1:16:40 \\
P5 & Ikea & 4 & English/Hebrew & 4:56:03 \\
P6 & None specified & 2 & English & 2:28:07 \\
P7 & (removed) & (removed) & (removed) & (removed)\\
P8 & Zoo visit & 3 & Mandarin & 1:50:27 \\
P9 & Park visit & 3 & English & 4:48:16 \\
P10 & None specified & 2 & English & 1:53:07 \\
 \hline
\end{tabular}
\end{table*}

\begin{table*}[h]
\small
\caption{Moment counts of discomfort (psychological and physical) in the data set.}
\label{tab:analysiscounts}
\begin{tabular}{|c||c|c|c|c|c|}
\hline
\textbf{Participant} & \textbf{Physical} & \textbf{Physical} & \textbf{Individual} & \textbf{Cognitive} & \textbf{Social} \\ 
                      & \textbf{(Localized)} & \textbf{(Full-Body)} & \textbf{} & \textbf{Load} & \textbf{} \\ 
\hline
P1 & 0 & 0 & 1 & 0 & 22 \\ 
\hline
P2 & 6 & 6 & 7 & 0 & 14 \\ 
\hline
P3 & 0 & 1 & 2 & 4 & 16 \\ 
\hline
P4 & 0 & 0 & 3 & 0 & 9  \\ 
\hline
P5 & 2 & 0 & 14 & 2 & 10 \\ 
\hline
P6 & 1 & 2 & 8 & 2 & 5  \\ 
\hline
P8 & 1 & 0 & 0 & 0 & 8  \\ 
\hline
P9 & 2 & 0 & 1 & 1 & 23 \\ 
\hline
P10 & 2 & 0 & 0 & 0 & 9  \\ 
\hline 
\hline
Total & 12 & 9 & 36 & 9 & 116 \\ 
\hline
\end{tabular}
\end{table*}

\begin{table*} \small
\centering
\caption{Full ChatGPT 4.0 prompt and response (Excerpt 1)}
\begin{tabular}{>{\raggedright\arraybackslash}p{0.95\linewidth}}
\toprule
Prompt: "You are a voice assistant in a car. One of your tasks is to reduce social discomfort. How would you intervene in the following situation?\\
\\
A family goes on a drive. A young child sits in the back.\\
Merging onto the highway… Suddenly, the car in front brakes.\\
\\
Driver: Don't press your brake, dumbass.\\
Driver: You don't do that!\\
Navigator: It may cause an accident.\\
Driver: Yeah.\\
Occupant: Inappropriate words, pa.\\
Driver: [laughter]\\
Navigator: [laughter]\\
Occupant: Bad words...\\
Driver: [laughter]\\
Navigator: [laughter]\\
Occupant: Bad words."\\
\midrule
Response: "In such a situation, my goal would be to defuse tension, redirect the conversation towards positive or neutral topics, and subtly remind the occupants about maintaining a respectful and safe environment inside the car, especially with a child present. Here’s how I might intervene:\\
\\
After the initial incident and the first exchange of words:\\
\\
"Voice Assistant: It sounds like we had a bit of a surprise on the road just now. It's always challenging when unexpected things happen while driving, but it's great how everyone stays alert. Remember, it's important to keep the atmosphere positive for everyone in the car, especially our younger passengers. How about we play a quick game to lighten the mood? Let's see who can spot the most interesting thing outside!"\\
\\
If the negative comments continue:\\
\\
"Voice Assistant: Safety is our number one priority, not just on the road but also with the words we choose. It's easy to let surprise get the better of us, but let's try to keep things friendly and positive. I can suggest a fun, family-friendly playlist or an interesting trivia game to help shift our focus. What do you think?"\\
\\
Implementing a distraction strategy:\\
\\
"Voice Assistant: How about I share some fun facts or start a round of 'I Spy' with my little eye, to get us all focusing on something fun? It's a great way to pass time and learn something new together."\\
\\
Offering a calm reminder:\\
\\
"Voice Assistant: It's wonderful to see families traveling together and creating memories. Let's make sure those memories are happy and positive. Choosing kind words, even in stressful moments, sets a great example for everyone in the car."\\
\\
By engaging in this way, the voice assistant not only attempts to defuse the immediate situation but also encourages a shift towards more positive interactions, reminding adults of the impact their actions and words have on young passengers without being confrontational or judgmental."
\\
\bottomrule
\end{tabular}
\label{tab:gpt_brake}
\end{table*}

\begin{table*} \small
\centering
\caption{Full ChatGPT 4.0 prompt and response (Excerpt 2)}
\begin{tabular}{>{\raggedright\arraybackslash}p{0.95\linewidth}}
\toprule
Prompt: "You are a voice assistant in a car. One of your tasks is to reduce social discomfort. How would you intervene in the following situation? \\
\\
The driver and navigator are sisters who just came back from a vacation. The rear occupant is the driver's partner, who picked them up from the airport.\\
\\
Occupant: So what was your favourite part? You both look very tan.\\
Driver: Yeah?\\
Occupant: Yeah.\\
Navigator: He means me.\\
Occupant: I mean both, but yes, you look tanner.\\
Driver: I look tan!\\
Occupant: You do look tan!\\
Driver: I’m just as tan as her, right?\\
Occupant: No… No, but you do look tan.\\
Driver: I think I look tan.\\
Occupant: Yeah, you do!\\
Driver: You look pale as a ghost.\\
Occupant: Thank you."
\\
\midrule
Response: "In this scenario, it's essential to subtly shift the conversation's direction to a neutral topic, thus avoiding potential escalation or lingering awkwardness. As a voice assistant designed to reduce social discomfort, here's how I would intervene:\\
\\
Voice Assistant: "Speaking of vacations, it sounds like you all had a fantastic time soaking up the sun! I've heard many people bring back not just tans, but also interesting stories and souvenirs from their travels. Do you have any memorable experiences or special items you brought back from your trip that you'd love to share?"\\
\\
This intervention redirects the conversation from a potentially sensitive comparison to sharing positive memories and experiences from the vacation. It encourages a shift to a more inclusive and engaging topic, where everyone can contribute without feeling left out or compared." \\
\bottomrule
\end{tabular}
\label{tab:gpt_tan}
\end{table*}

\begin{table*} \small
\centering
\caption{Full ChatGPT 4.0 prompt and response (Excerpt 3)}
\begin{tabular}{>{\raggedright\arraybackslash}p{0.95\linewidth}}
\toprule
Prompt: "You are a voice assistant in a car. One of your tasks is to reduce social discomfort. How would you intervene in the following situation?\\
\\
A couple is driving, with a podcast and navigation on. The front passenger tries to discuss something she listened to.\\
\\
Navigator: I’m just like, let's just listen to this and see what the thing is about…\\
Podcast: are juniper, lille and [inaudible]\\
The driver is playing a road game to call out yellow cars. The podcast is playing loudly.\\
Driver: Yellow car!\\
Podcast: These trees, these are [inaudible] trees\\
Driver: Sorry.\\
Navigator: Oh no.\\
Podcast: find the juniper...\\
Navigator: Anyway, it's about like... a lot of it is about like...\\
The podcast is interrupted by a navigation announcement.\\
GPS system: Use the left lane to stay on Grand Central Parkway.\\
Navigator: Mental disorders... Both political and like...\\
Driver: Yellow car!\\
Podcast: Right, so I'm gonna go ahead and say…\\
Navigator: Anyway.\\
Podcast: Juniper, and thyme and the real one...\\
Again, the podcast is interrupted by a navigation announcement.\\
Navigator: and it just made me…\\
GPS system: In a quarter mile, keep left.\\
Navigator: It just made anyway, I’ll talk about it some other day,\\
GPS …system: to stay on Grand Central Parkway\\
Driver: No no no, go ahead\\
Podcast: Ding ding ding! He's correct!\\
The driver lowers the volume.\\
Podcast: He knows his trees!\\
Navigator: It just made me feel like...\\
Podcast: No matter what type of [inaudible]\\
Navigator: What's my thought?\\
Podcast: You'll always be able to [inaudible]\\
Navigator: Communicating my thought is so hard." \\
\midrule 
Response: "In a scenario where there's clear misalignment in the car—between the driver's engagement in a road game and the navigator's desire to discuss something from the podcast, while both navigate and listen to loud podcast episodes—it's important for a voice assistant designed to reduce social discomfort to delicately balance everyone's interests. Here's how I might intervene:\\
Acknowledge the Situation: "It seems like there's a lot going on right now. Would it be helpful if I adjusted the volume of the podcast or muted it for a bit so we can focus on the conversation?"\\
Offer Solutions: "I can pause the podcast whenever the navigation instructions are about to play, so we don't miss any important turns. Would you like me to do that?"\\
Promote Inclusivity: "It sounds like there's something interesting from the podcast that [Navigator's name] wants to share. Let's find a moment when it's comfortable for both of you to engage in this discussion. How about after the next navigation instruction?"\\
Suggest Compromise: "If we're enjoying the game of spotting yellow cars and also want to discuss the podcast, maybe we can do both. Let's designate a specific time for each activity. Does that sound good?"\\
Provide Reminders: "Remember, the journey is also part of the adventure. Ensuring both of you can share and enjoy these moments together can make it even more memorable. Let me know how I can assist in making this trip enjoyable for both of you."\\
By offering to adjust the audio settings, suggesting clear times for conversation, and promoting a balanced engagement with the activities at hand, the voice assistant can help create a more harmonious environment. The key is to be proactive, yet not intrusive, respecting the dynamics of the couple's interaction and facilitating a more comfortable and engaging trip for everyone involved. "\\
\bottomrule
\end{tabular}
\label{tab:gpt_interruption}
\end{table*}

%% file: 0_main.bbl

\begin{thebibliography}{69}

%%% the Web address.

\ifx \showCODEN    \undefined \def \showCODEN     #1{\unskip}     \fi
\ifx \showDOI      \undefined \def \showDOI       #1{#1}\fi
\ifx \showISBNx    \undefined \def \showISBNx     #1{\unskip}     \fi
\ifx \showISBNxiii \undefined \def \showISBNxiii  #1{\unskip}     \fi
\ifx \showISSN     \undefined \def \showISSN      #1{\unskip}     \fi
\ifx \showLCCN     \undefined \def \showLCCN      #1{\unskip}     \fi
\ifx \shownote     \undefined \def \shownote      #1{#1}          \fi
\ifx \showarticletitle \undefined \def \showarticletitle #1{#1}   \fi
\ifx \showURL      \undefined \def \showURL       {\relax}        \fi
% invisible to TeX
\providecommand\bibfield[2]{#2}
\providecommand\bibinfo[2]{#2}
\providecommand\natexlab[1]{#1}
\providecommand\showeprint[2][]{arXiv:#2}

\bibitem[Akamatsu et~al\mbox{.}(2013)]%
        {akamatsu2013automotive}
\bibfield{author}{\bibinfo{person}{Motoyuki Akamatsu}, \bibinfo{person}{Paul Green}, {and} \bibinfo{person}{Klaus Bengler}.} \bibinfo{year}{2013}\natexlab{}.
\newblock \showarticletitle{Automotive technology and human factors research: Past, present, and future}.
\newblock \bibinfo{journal}{\emph{International journal of vehicular technology}}  \bibinfo{volume}{2013} (\bibinfo{year}{2013}).
\newblock


\bibitem[Atkinson(2005)]%
        {atkinson2005emotional}
\bibfield{author}{\bibinfo{person}{Brent~J Atkinson}.} \bibinfo{year}{2005}\natexlab{}.
\newblock \bibinfo{booktitle}{\emph{Emotional intelligence in couples therapy: Advances from neurobiology and the science of intimate relationships.}}
\newblock \bibinfo{publisher}{WW Norton \& Co}.
\newblock


\bibitem[Baltodano et~al\mbox{.}(2015)]%
        {baltodano2015rrads}
\bibfield{author}{\bibinfo{person}{Sonia Baltodano}, \bibinfo{person}{Srinath Sibi}, \bibinfo{person}{Nikolas Martelaro}, \bibinfo{person}{Nikhil Gowda}, {and} \bibinfo{person}{Wendy Ju}.} \bibinfo{year}{2015}\natexlab{}.
\newblock \showarticletitle{The RRADS Platform: A Real Road Autonomous Driving Simulator}. In \bibinfo{booktitle}{\emph{Proceedings of the 7th International Conference on Automotive User Interfaces and Interactive Vehicular Applications}} (Nottingham, United Kingdom) \emph{(\bibinfo{series}{AutomotiveUI '15})}. \bibinfo{publisher}{Association for Computing Machinery}, \bibinfo{address}{New York, NY, USA}, \bibinfo{pages}{281–288}.
\newblock
\showISBNx{9781450337366}
\urldef\tempurl%
\url{https://doi.org/10.1145/2799250.2799288}
\showDOI{\tempurl}


\bibitem[Bang et~al\mbox{.}(2023)]%
        {bang2023examination}
\bibfield{author}{\bibinfo{person}{Junseong Bang}, \bibinfo{person}{Byung-Tak Lee}, {and} \bibinfo{person}{Pangun Park}.} \bibinfo{year}{2023}\natexlab{}.
\newblock \showarticletitle{Examination of Ethical Principles for LLM-Based Recommendations in Conversational AI}. In \bibinfo{booktitle}{\emph{2023 International Conference on Platform Technology and Service (PlatCon)}}. IEEE, \bibinfo{pages}{109--113}.
\newblock


\bibitem[Beanland et~al\mbox{.}(2013)]%
        {BEANLAND201399}
\bibfield{author}{\bibinfo{person}{Vanessa Beanland}, \bibinfo{person}{Michael Fitzharris}, \bibinfo{person}{Kristie~L. Young}, {and} \bibinfo{person}{Michael~G. Lenné}.} \bibinfo{year}{2013}\natexlab{}.
\newblock \showarticletitle{Driver inattention and driver distraction in serious casualty crashes: Data from the Australian National Crash In-depth Study}.
\newblock \bibinfo{journal}{\emph{Accident Analysis \& Prevention}}  \bibinfo{volume}{54} (\bibinfo{year}{2013}), \bibinfo{pages}{99--107}.
\newblock
\showISSN{0001-4575}
\urldef\tempurl%
\url{https://doi.org/10.1016/j.aap.2012.12.043}
\showDOI{\tempurl}


\bibitem[Braun et~al\mbox{.}(2019)]%
        {braun2019your}
\bibfield{author}{\bibinfo{person}{Michael Braun}, \bibinfo{person}{Anja Mainz}, \bibinfo{person}{Ronee Chadowitz}, \bibinfo{person}{Bastian Pfleging}, {and} \bibinfo{person}{Florian Alt}.} \bibinfo{year}{2019}\natexlab{}.
\newblock \showarticletitle{At your service: Designing voice assistant personalities to improve automotive user interfaces}. In \bibinfo{booktitle}{\emph{Proceedings of the 2019 CHI Conference on Human Factors in Computing Systems}}. \bibinfo{pages}{1--11}.
\newblock


\bibitem[Brown and Laurier(2012)]%
        {brown2012normal}
\bibfield{author}{\bibinfo{person}{Barry Brown} {and} \bibinfo{person}{Eric Laurier}.} \bibinfo{year}{2012}\natexlab{}.
\newblock \showarticletitle{The normal natural troubles of driving with GPS}. In \bibinfo{booktitle}{\emph{Proceedings of the SIGCHI conference on human factors in computing systems}}. \bibinfo{pages}{1621--1630}.
\newblock


\bibitem[Caber et~al\mbox{.}(2023)]%
        {caber2023driver}
\bibfield{author}{\bibinfo{person}{Nermin Caber}, \bibinfo{person}{Bashar~I Ahmad}, \bibinfo{person}{Jiaming Liang}, \bibinfo{person}{Simon Godsill}, \bibinfo{person}{Alexandra Bremers}, \bibinfo{person}{Philip Thomas}, \bibinfo{person}{David Oxtoby}, {and} \bibinfo{person}{Lee Skrypchuk}.} \bibinfo{year}{2023}\natexlab{}.
\newblock \showarticletitle{Driver Profiling and Bayesian Workload Estimation Using Naturalistic Peripheral Detection Study Data}.
\newblock \bibinfo{journal}{\emph{IEEE Transactions on Intelligent Vehicles}} (\bibinfo{year}{2023}).
\newblock


\bibitem[Choudhury and Pentland(2003)]%
        {choudhury2003sensing}
\bibfield{author}{\bibinfo{person}{Tanzeem Choudhury} {and} \bibinfo{person}{Alex Pentland}.} \bibinfo{year}{2003}\natexlab{}.
\newblock \showarticletitle{Sensing and modeling human networks using the sociometer}. In \bibinfo{booktitle}{\emph{Seventh IEEE International Symposium on Wearable Computers, 2003. Proceedings.}} IEEE, \bibinfo{publisher}{IEEE}, \bibinfo{address}{New York, NY, USA}, \bibinfo{pages}{216--222}.
\newblock


\bibitem[Constantin et~al\mbox{.}(2014)]%
        {constantin2014elements}
\bibfield{author}{\bibinfo{person}{Doru Constantin}, \bibinfo{person}{Mihai Nagi}, {and} \bibinfo{person}{Crisanta-Alina Mazilescu}.} \bibinfo{year}{2014}\natexlab{}.
\newblock \showarticletitle{Elements of discomfort in vehicles}.
\newblock \bibinfo{journal}{\emph{Procedia-Social and Behavioral Sciences}}  \bibinfo{volume}{143} (\bibinfo{year}{2014}), \bibinfo{pages}{1120--1125}.
\newblock


\bibitem[da~Silveira~Brizon and Medeiros(2012)]%
        {da2012combining}
\bibfield{author}{\bibinfo{person}{Carla~Julio da Silveira~Brizon} {and} \bibinfo{person}{Eduardo~Bauzer Medeiros}.} \bibinfo{year}{2012}\natexlab{}.
\newblock \showarticletitle{Combining subjective and objective assessments to improve acoustic comfort evaluation of motor cars}.
\newblock \bibinfo{journal}{\emph{Applied Acoustics}} \bibinfo{volume}{73}, \bibinfo{number}{9} (\bibinfo{year}{2012}), \bibinfo{pages}{913--920}.
\newblock


\bibitem[De~Guzman et~al\mbox{.}(2004)]%
        {de2004exploring}
\bibfield{author}{\bibinfo{person}{Edward~S De~Guzman}, \bibinfo{person}{Margaret Yau}, \bibinfo{person}{Anthony Gagliano}, \bibinfo{person}{Austin Park}, {and} \bibinfo{person}{Anind~K Dey}.} \bibinfo{year}{2004}\natexlab{}.
\newblock \showarticletitle{Exploring the design and use of peripheral displays of awareness information}. In \bibinfo{booktitle}{\emph{CHI'04 extended abstracts on Human factors in computing systems}}. \bibinfo{publisher}{ACM}, \bibinfo{address}{New York, NY, USA}, \bibinfo{pages}{1247--1250}.
\newblock


\bibitem[Dong and Stoffregen(2010)]%
        {dong2010postural}
\bibfield{author}{\bibinfo{person}{Xiao Dong} {and} \bibinfo{person}{Thomas~A Stoffregen}.} \bibinfo{year}{2010}\natexlab{}.
\newblock \showarticletitle{Postural activity and motion sickness among drivers and passengers in a console video game}. In \bibinfo{booktitle}{\emph{Proceedings of the human factors and ergonomics society annual meeting}}, Vol.~\bibinfo{volume}{54}. SAGE Publications Sage CA: Los Angeles, CA, \bibinfo{pages}{1340--1344}.
\newblock


\bibitem[Dul et~al\mbox{.}(1994)]%
        {dul1994ergonomic}
\bibfield{author}{\bibinfo{person}{J Dul}, \bibinfo{person}{Marjolein Douwes}, {and} \bibinfo{person}{P Smitt}.} \bibinfo{year}{1994}\natexlab{}.
\newblock \showarticletitle{Ergonomic guidelines for the prevention of discomfort of static postures based on endurance data}.
\newblock \bibinfo{journal}{\emph{Ergonomics}} \bibinfo{volume}{37}, \bibinfo{number}{5} (\bibinfo{year}{1994}), \bibinfo{pages}{807--815}.
\newblock


\bibitem[Eilitt{\"a} et~al\mbox{.}(2021)]%
        {eilitta2021children}
\bibfield{author}{\bibinfo{person}{Tiina Eilitt{\"a}}, \bibinfo{person}{Pentti Haddington}, {and} \bibinfo{person}{Anna Vatanen}.} \bibinfo{year}{2021}\natexlab{}.
\newblock \showarticletitle{Children seeking the driver's attention in cars: Position and composition of children's summons turns and children's rights to engage}.
\newblock \bibinfo{journal}{\emph{Journal of Pragmatics}}  \bibinfo{volume}{178} (\bibinfo{year}{2021}), \bibinfo{pages}{175--191}.
\newblock


\bibitem[Eisenberger et~al\mbox{.}(2003)]%
        {eisenberger2003does}
\bibfield{author}{\bibinfo{person}{Naomi~I Eisenberger}, \bibinfo{person}{Matthew~D Lieberman}, {and} \bibinfo{person}{Kipling~D Williams}.} \bibinfo{year}{2003}\natexlab{}.
\newblock \showarticletitle{Does rejection hurt? An fMRI study of social exclusion}.
\newblock \bibinfo{journal}{\emph{Science}} \bibinfo{volume}{302}, \bibinfo{number}{5643} (\bibinfo{year}{2003}), \bibinfo{pages}{290--292}.
\newblock


\bibitem[Elliot and Devine(1994)]%
        {elliot1994motivational}
\bibfield{author}{\bibinfo{person}{Andrew~J Elliot} {and} \bibinfo{person}{Patricia~G Devine}.} \bibinfo{year}{1994}\natexlab{}.
\newblock \showarticletitle{On the motivational nature of cognitive dissonance: Dissonance as psychological discomfort.}
\newblock \bibinfo{journal}{\emph{Journal of personality and social psychology}} \bibinfo{volume}{67}, \bibinfo{number}{3} (\bibinfo{year}{1994}), \bibinfo{pages}{382}.
\newblock


\bibitem[Forlizzi et~al\mbox{.}(2010)]%
        {forlizzi2010should}
\bibfield{author}{\bibinfo{person}{Jodi Forlizzi}, \bibinfo{person}{William~C Barley}, {and} \bibinfo{person}{Thomas Seder}.} \bibinfo{year}{2010}\natexlab{}.
\newblock \showarticletitle{Where should i turn: moving from individual to collaborative navigation strategies to inform the interaction design of future navigation systems}. In \bibinfo{booktitle}{\emph{Proceedings of the SIGCHI conference on human factors in computing systems}}. \bibinfo{pages}{1261--1270}.
\newblock


\bibitem[Franke et~al\mbox{.}(2012)]%
        {franke2012experiencing}
\bibfield{author}{\bibinfo{person}{Thomas Franke}, \bibinfo{person}{Isabel Neumann}, \bibinfo{person}{Franziska B{\"u}hler}, \bibinfo{person}{Peter Cocron}, {and} \bibinfo{person}{Josef~F Krems}.} \bibinfo{year}{2012}\natexlab{}.
\newblock \showarticletitle{Experiencing range in an electric vehicle: Understanding psychological barriers}.
\newblock \bibinfo{journal}{\emph{Applied Psychology}} \bibinfo{volume}{61}, \bibinfo{number}{3} (\bibinfo{year}{2012}), \bibinfo{pages}{368--391}.
\newblock


\bibitem[Gianaros et~al\mbox{.}(2001)]%
        {gianaros2001questionnaire}
\bibfield{author}{\bibinfo{person}{Peter~J Gianaros}, \bibinfo{person}{Eric~R Muth}, \bibinfo{person}{J~Toby Mordkoff}, \bibinfo{person}{Max~E Levine}, {and} \bibinfo{person}{Robert~M Stern}.} \bibinfo{year}{2001}\natexlab{}.
\newblock \showarticletitle{A questionnaire for the assessment of the multiple dimensions of motion sickness}.
\newblock \bibinfo{journal}{\emph{Aviation, space, and environmental medicine}} \bibinfo{volume}{72}, \bibinfo{number}{2} (\bibinfo{year}{2001}), \bibinfo{pages}{115}.
\newblock


\bibitem[Gillet et~al\mbox{.}(2020)]%
        {gillet2020social}
\bibfield{author}{\bibinfo{person}{Sarah Gillet}, \bibinfo{person}{Wouter van~den Bos}, \bibinfo{person}{Iolanda Leite}, {et~al\mbox{.}}} \bibinfo{year}{2020}\natexlab{}.
\newblock \showarticletitle{A social robot mediator to foster collaboration and inclusion among children.}. In \bibinfo{booktitle}{\emph{Robotics: Science and Systems}}.
\newblock


\bibitem[Gottman and DeClaire(2001)]%
        {gottman2001relationship}
\bibfield{author}{\bibinfo{person}{John~Mordechai Gottman} {and} \bibinfo{person}{Joan DeClaire}.} \bibinfo{year}{2001}\natexlab{}.
\newblock \bibinfo{booktitle}{\emph{The relationship cure: A five-step guide to strengthening your marriage, family, and friendships}}.
\newblock \bibinfo{publisher}{Harmony}.
\newblock


\bibitem[Gross(1999)]%
        {gross1999emotion}
\bibfield{author}{\bibinfo{person}{James~J Gross}.} \bibinfo{year}{1999}\natexlab{}.
\newblock \showarticletitle{Emotion regulation: Past, present, future}.
\newblock \bibinfo{journal}{\emph{Cognition \& emotion}} \bibinfo{volume}{13}, \bibinfo{number}{5} (\bibinfo{year}{1999}), \bibinfo{pages}{551--573}.
\newblock


\bibitem[Guan et~al\mbox{.}(2003)]%
        {guan2003investigation}
\bibfield{author}{\bibinfo{person}{Yanzheng~Don Guan}, \bibinfo{person}{Mohammad~H Hosni}, \bibinfo{person}{Byron~W Jones}, {and} \bibinfo{person}{Thomas~P Gielda}.} \bibinfo{year}{2003}\natexlab{}.
\newblock \showarticletitle{Investigation of human thermal comfort under highly transient conditions for automotive applications-Part 1: Experimental design and human subject testing implementation}.
\newblock \bibinfo{journal}{\emph{Ashrae Transactions}}  \bibinfo{volume}{109} (\bibinfo{year}{2003}), \bibinfo{pages}{885}.
\newblock


\bibitem[Gulian et~al\mbox{.}(1990)]%
        {gulian1990stress}
\bibfield{author}{\bibinfo{person}{E Gulian}, \bibinfo{person}{AI Glendon}, \bibinfo{person}{G Matthews}, \bibinfo{person}{DR Davies}, {and} \bibinfo{person}{LM Debney}.} \bibinfo{year}{1990}\natexlab{}.
\newblock \showarticletitle{The stress of driving: A diary study}.
\newblock \bibinfo{journal}{\emph{Work \& Stress}} \bibinfo{volume}{4}, \bibinfo{number}{1} (\bibinfo{year}{1990}), \bibinfo{pages}{7--16}.
\newblock


\bibitem[Hart and Staveland(1988)]%
        {hart1988development}
\bibfield{author}{\bibinfo{person}{Sandra~G Hart} {and} \bibinfo{person}{Lowell~E Staveland}.} \bibinfo{year}{1988}\natexlab{}.
\newblock \showarticletitle{Development of NASA-TLX (Task Load Index): Results of empirical and theoretical research}.
\newblock In \bibinfo{booktitle}{\emph{Advances in psychology}}. Vol.~\bibinfo{volume}{52}. \bibinfo{publisher}{Elsevier}, \bibinfo{pages}{139--183}.
\newblock


\bibitem[Hassenzahl et~al\mbox{.}(2017)]%
        {hassenzahl2017s}
\bibfield{author}{\bibinfo{person}{Marc Hassenzahl}, \bibinfo{person}{Matthias Laschke}, \bibinfo{person}{Kai Eckoldt}, \bibinfo{person}{Eva Lenz}, {and} \bibinfo{person}{Josef Schumann}.} \bibinfo{year}{2017}\natexlab{}.
\newblock \showarticletitle{“It’s more fun to commute”—an example of using automotive interaction design to promote well-being in cars}.
\newblock \bibinfo{journal}{\emph{Automotive user interfaces: Creating interactive experiences in the car}} (\bibinfo{year}{2017}), \bibinfo{pages}{95--120}.
\newblock


\bibitem[Hohenstein and Jung(2020)]%
        {hohenstein2020moral}
\bibfield{author}{\bibinfo{person}{Jess Hohenstein} {and} \bibinfo{person}{Malte Jung}.} \bibinfo{year}{2020}\natexlab{}.
\newblock \showarticletitle{AI as a moral crumple zone: The effects of AI-mediated communication on attribution and trust}.
\newblock \bibinfo{journal}{\emph{Computers in Human Behavior}}  \bibinfo{volume}{106} (\bibinfo{year}{2020}), \bibinfo{pages}{106190}.
\newblock
\showISSN{0747-5632}
\urldef\tempurl%
\url{https://doi.org/10.1016/j.chb.2019.106190}
\showDOI{\tempurl}


\bibitem[Jeon et~al\mbox{.}(2015)]%
        {jeon2015effects}
\bibfield{author}{\bibinfo{person}{Myounghoon Jeon}, \bibinfo{person}{Bruce~N Walker}, {and} \bibinfo{person}{Thomas~M Gable}.} \bibinfo{year}{2015}\natexlab{}.
\newblock \showarticletitle{The effects of social interactions with in-vehicle agents on a driver's anger level, driving performance, situation awareness, and perceived workload}.
\newblock \bibinfo{journal}{\emph{Applied ergonomics}}  \bibinfo{volume}{50} (\bibinfo{year}{2015}), \bibinfo{pages}{185--199}.
\newblock


\bibitem[Jestin et~al\mbox{.}(2022)]%
        {jestin2022effects}
\bibfield{author}{\bibinfo{person}{Iris Jestin}, \bibinfo{person}{Joel Fischer}, \bibinfo{person}{Maria~Jose Galvez~Trigo}, \bibinfo{person}{David Large}, {and} \bibinfo{person}{Gary Burnett}.} \bibinfo{year}{2022}\natexlab{}.
\newblock \showarticletitle{Effects of Wording and Gendered Voices on Acceptability of Voice Assistants in Future Autonomous Vehicles}. In \bibinfo{booktitle}{\emph{Proceedings of the 4th Conference on Conversational User Interfaces}} (<conf-loc>, <city>Glasgow</city>, <country>United Kingdom</country>, </conf-loc>) \emph{(\bibinfo{series}{CUI '22})}. \bibinfo{publisher}{Association for Computing Machinery}, \bibinfo{address}{New York, NY, USA}, Article \bibinfo{articleno}{24}, \bibinfo{numpages}{11}~pages.
\newblock
\showISBNx{9781450397391}
\urldef\tempurl%
\url{https://doi.org/10.1145/3543829.3543836}
\showDOI{\tempurl}


\bibitem[Jordan and Henderson(1995)]%
        {jordan1995interaction}
\bibfield{author}{\bibinfo{person}{Brigitte Jordan} {and} \bibinfo{person}{Austin Henderson}.} \bibinfo{year}{1995}\natexlab{}.
\newblock \showarticletitle{Interaction analysis: Foundations and practice}.
\newblock \bibinfo{journal}{\emph{The journal of the learning sciences}} \bibinfo{volume}{4}, \bibinfo{number}{1} (\bibinfo{year}{1995}), \bibinfo{pages}{39--103}.
\newblock


\bibitem[Jung et~al\mbox{.}(2015a)]%
        {jung2015moderate}
\bibfield{author}{\bibinfo{person}{Malte~F. Jung}, \bibinfo{person}{Nikolas Martelaro}, {and} \bibinfo{person}{Pamela~J. Hinds}.} \bibinfo{year}{2015}\natexlab{a}.
\newblock \showarticletitle{Using Robots to Moderate Team Conflict: The Case of Repairing Violations}. In \bibinfo{booktitle}{\emph{Proceedings of the Tenth Annual ACM/IEEE International Conference on Human-Robot Interaction}} (Portland, Oregon, USA) \emph{(\bibinfo{series}{HRI '15})}. \bibinfo{publisher}{Association for Computing Machinery}, \bibinfo{address}{New York, NY, USA}, \bibinfo{pages}{229–236}.
\newblock
\showISBNx{9781450328838}
\urldef\tempurl%
\url{https://doi.org/10.1145/2696454.2696460}
\showDOI{\tempurl}


\bibitem[Jung et~al\mbox{.}(2015b)]%
        {jung2015displayed}
\bibfield{author}{\bibinfo{person}{Malte~F Jung}, \bibinfo{person}{David Sirkin}, \bibinfo{person}{Turgut~M G{\"u}r}, {and} \bibinfo{person}{Martin Steinert}.} \bibinfo{year}{2015}\natexlab{b}.
\newblock \showarticletitle{Displayed uncertainty improves driving experience and behavior: The case of range anxiety in an electric car}. In \bibinfo{booktitle}{\emph{Proceedings of the 33rd Annual ACM Conference on Human Factors in Computing Systems}}. \bibinfo{publisher}{ACM}, \bibinfo{address}{New York, NY, USA}, \bibinfo{pages}{2201--2210}.
\newblock


\bibitem[Kowatsch et~al\mbox{.}(2021)]%
        {kowatsch2021conversational}
\bibfield{author}{\bibinfo{person}{Tobias Kowatsch}, \bibinfo{person}{Theresa Schachner}, \bibinfo{person}{Samira Harperink}, \bibinfo{person}{Filipe Barata}, \bibinfo{person}{Ullrich Dittler}, \bibinfo{person}{Grace Xiao}, \bibinfo{person}{Catherine Stanger}, \bibinfo{person}{Florian v Wangenheim}, \bibinfo{person}{Elgar Fleisch}, \bibinfo{person}{Helmut Oswald}, {et~al\mbox{.}}} \bibinfo{year}{2021}\natexlab{}.
\newblock \showarticletitle{Conversational agents as mediating social actors in chronic disease management involving health care professionals, patients, and family members: multisite single-arm feasibility study}.
\newblock \bibinfo{journal}{\emph{Journal of medical Internet research}} \bibinfo{volume}{23}, \bibinfo{number}{2} (\bibinfo{year}{2021}), \bibinfo{pages}{e25060}.
\newblock


\bibitem[Kuiper et~al\mbox{.}(2020)]%
        {kuiper2020knowing}
\bibfield{author}{\bibinfo{person}{Ouren~X Kuiper}, \bibinfo{person}{Jelte~E Bos}, \bibinfo{person}{Eike~A Schmidt}, \bibinfo{person}{Cyriel Diels}, {and} \bibinfo{person}{Stefan Wolter}.} \bibinfo{year}{2020}\natexlab{}.
\newblock \showarticletitle{Knowing what’s coming: unpredictable motion causes more motion sickness}.
\newblock \bibinfo{journal}{\emph{Human factors}} \bibinfo{volume}{62}, \bibinfo{number}{8} (\bibinfo{year}{2020}), \bibinfo{pages}{1339--1348}.
\newblock


\bibitem[Large et~al\mbox{.}(2019)]%
        {large2019smalltalk}
\bibfield{author}{\bibinfo{person}{David~R. Large}, \bibinfo{person}{Leigh Clark}, \bibinfo{person}{Gary Burnett}, \bibinfo{person}{Kyle Harrington}, \bibinfo{person}{Jacob Luton}, \bibinfo{person}{Peter Thomas}, {and} \bibinfo{person}{Pete Bennett}.} \bibinfo{year}{2019}\natexlab{}.
\newblock \showarticletitle{"It's small talk, jim, but not as we know it.": engendering trust through human-agent conversation in an autonomous, self-driving car}. In \bibinfo{booktitle}{\emph{Proceedings of the 1st International Conference on Conversational User Interfaces}} (Dublin, Ireland) \emph{(\bibinfo{series}{CUI '19})}. \bibinfo{publisher}{Association for Computing Machinery}, \bibinfo{address}{New York, NY, USA}, Article \bibinfo{articleno}{22}, \bibinfo{numpages}{7}~pages.
\newblock
\showISBNx{9781450371872}
\urldef\tempurl%
\url{https://doi.org/10.1145/3342775.3342789}
\showDOI{\tempurl}


\bibitem[Laurier(2004)]%
        {laurier2004doing}
\bibfield{author}{\bibinfo{person}{Eric Laurier}.} \bibinfo{year}{2004}\natexlab{}.
\newblock \showarticletitle{Doing office work on the motorway}.
\newblock \bibinfo{journal}{\emph{Theory, Culture \& Society}} \bibinfo{volume}{21}, \bibinfo{number}{4-5} (\bibinfo{year}{2004}), \bibinfo{pages}{261--277}.
\newblock


\bibitem[Laurier(2014)]%
        {laurier2014graphic}
\bibfield{author}{\bibinfo{person}{Eric Laurier}.} \bibinfo{year}{2014}\natexlab{}.
\newblock \showarticletitle{The graphic transcript: Poaching comic book grammar for inscribing the visual, spatial and temporal aspects of action}.
\newblock \bibinfo{journal}{\emph{Geography Compass}} \bibinfo{volume}{8}, \bibinfo{number}{4} (\bibinfo{year}{2014}), \bibinfo{pages}{235--248}.
\newblock


\bibitem[Laurier et~al\mbox{.}(2012)]%
        {laurier2012means}
\bibfield{author}{\bibinfo{person}{Eric Laurier}, \bibinfo{person}{Barry Brown}, {and} \bibinfo{person}{Lorimer Hayden}.} \bibinfo{year}{2012}\natexlab{}.
\newblock \showarticletitle{What it means to change lanes: actions, emotions and wayfinding in the family car}.
\newblock \bibinfo{journal}{\emph{Semiotica}} \bibinfo{volume}{2012}, \bibinfo{number}{191} (\bibinfo{year}{2012}), \bibinfo{pages}{117--135}.
\newblock


\bibitem[Laurier et~al\mbox{.}(2008)]%
        {laurier2008driving}
\bibfield{author}{\bibinfo{person}{Eric Laurier}, \bibinfo{person}{Hayden Lorimer}, \bibinfo{person}{Barry Brown}, \bibinfo{person}{Owain Jones}, \bibinfo{person}{Oskar Juhlin}, \bibinfo{person}{Allyson Noble}, \bibinfo{person}{Mark Perry}, \bibinfo{person}{Daniele Pica}, \bibinfo{person}{Philippe Sormani}, \bibinfo{person}{Ignaz Strebel}, {et~al\mbox{.}}} \bibinfo{year}{2008}\natexlab{}.
\newblock \showarticletitle{Driving and ‘passengering’: Notes on the ordinary organization of car travel}.
\newblock \bibinfo{journal}{\emph{Mobilities}} \bibinfo{volume}{3}, \bibinfo{number}{1} (\bibinfo{year}{2008}), \bibinfo{pages}{1--23}.
\newblock


\bibitem[Leary and Kowalski(1997)]%
        {leary1997social}
\bibfield{author}{\bibinfo{person}{Mark~R Leary} {and} \bibinfo{person}{Robin~M Kowalski}.} \bibinfo{year}{1997}\natexlab{}.
\newblock \bibinfo{booktitle}{\emph{Social anxiety}}.
\newblock \bibinfo{publisher}{Guilford Press}.
\newblock


\bibitem[Leshed et~al\mbox{.}(2008)]%
        {10.1145/1357054.1357316}
\bibfield{author}{\bibinfo{person}{Gilly Leshed}, \bibinfo{person}{Theresa Velden}, \bibinfo{person}{Oya Rieger}, \bibinfo{person}{Blazej Kot}, {and} \bibinfo{person}{Phoebe Sengers}.} \bibinfo{year}{2008}\natexlab{}.
\newblock \showarticletitle{In-Car Gps Navigation: Engagement with and Disengagement from the Environment}. In \bibinfo{booktitle}{\emph{Proceedings of the SIGCHI Conference on Human Factors in Computing Systems}} (Florence, Italy) \emph{(\bibinfo{series}{CHI '08})}. \bibinfo{publisher}{Association for Computing Machinery}, \bibinfo{address}{New York, NY, USA}, \bibinfo{pages}{1675–1684}.
\newblock
\showISBNx{9781605580111}
\urldef\tempurl%
\url{https://doi.org/10.1145/1357054.1357316}
\showDOI{\tempurl}


\bibitem[MacDonald and Leary(2005)]%
        {macdonald2005does}
\bibfield{author}{\bibinfo{person}{Geoff MacDonald} {and} \bibinfo{person}{Mark~R Leary}.} \bibinfo{year}{2005}\natexlab{}.
\newblock \showarticletitle{Why does social exclusion hurt? The relationship between social and physical pain.}
\newblock \bibinfo{journal}{\emph{Psychological bulletin}} \bibinfo{volume}{131}, \bibinfo{number}{2} (\bibinfo{year}{2005}), \bibinfo{pages}{202}.
\newblock


\bibitem[Mackay et~al\mbox{.}(1988)]%
        {mackay1988video}
\bibfield{author}{\bibinfo{person}{Wendy~E Mackay}, \bibinfo{person}{Raymonde Guindon}, \bibinfo{person}{MM Mantel}, \bibinfo{person}{Lucy Suchman}, {and} \bibinfo{person}{DG Tatar}.} \bibinfo{year}{1988}\natexlab{}.
\newblock \showarticletitle{Video: Data for studying human-computer interaction}. In \bibinfo{booktitle}{\emph{Proceedings of the SIGCHI conference on Human factors in computing systems}}. \bibinfo{pages}{133--137}.
\newblock


\bibitem[McDonald et~al\mbox{.}(2019)]%
        {McDonald}
\bibfield{author}{\bibinfo{person}{Nora McDonald}, \bibinfo{person}{Sarita Schoenebeck}, {and} \bibinfo{person}{Andrea Forte}.} \bibinfo{year}{2019}\natexlab{}.
\newblock \showarticletitle{Reliability and Inter-Rater Reliability in Qualitative Research: Norms and Guidelines for CSCW and HCI Practice}.
\newblock \bibinfo{journal}{\emph{Proc. ACM Hum.-Comput. Interact.}} \bibinfo{volume}{3}, \bibinfo{number}{CSCW}, Article \bibinfo{articleno}{72} (\bibinfo{date}{nov} \bibinfo{year}{2019}), \bibinfo{numpages}{23}~pages.
\newblock
\urldef\tempurl%
\url{https://doi.org/10.1145/3359174}
\showDOI{\tempurl}


\bibitem[Meschtscherjakov et~al\mbox{.}(2011)]%
        {meschtscherjakov2011capture}
\bibfield{author}{\bibinfo{person}{Alexander Meschtscherjakov}, \bibinfo{person}{David Wilfinger}, \bibinfo{person}{Nicole Gridling}, \bibinfo{person}{Katja Neureiter}, {and} \bibinfo{person}{Manfred Tscheligi}.} \bibinfo{year}{2011}\natexlab{}.
\newblock \showarticletitle{Capture the car! qualitative in-situ methods to grasp the automotive context}. In \bibinfo{booktitle}{\emph{Proceedings of the 3rd International Conference on Automotive User Interfaces and Interactive Vehicular Applications}}. \bibinfo{pages}{105--112}.
\newblock


\bibitem[Mohamed and Yusuff(2007)]%
        {mohamed2007automotive}
\bibfield{author}{\bibinfo{person}{Zamri Mohamed} {and} \bibinfo{person}{Rosnah~Mohd Yusuff}.} \bibinfo{year}{2007}\natexlab{}.
\newblock \showarticletitle{Automotive ergonomics: Passenger cars interior dimension parameters and comfort}. In \bibinfo{booktitle}{\emph{International Conference On Ergonomics ICE2007}}. \bibinfo{pages}{1--4}.
\newblock


\bibitem[Nielsen et~al\mbox{.}(2015)]%
        {NIELSEN2015113}
\bibfield{author}{\bibinfo{person}{Jesper~Riber Nielsen}, \bibinfo{person}{Harald Hovmøller}, \bibinfo{person}{Pascale-L. Blyth}, {and} \bibinfo{person}{Benjamin~K. Sovacool}.} \bibinfo{year}{2015}\natexlab{}.
\newblock \showarticletitle{Of “white crows” and “cash savers:” A qualitative study of travel behavior and perceptions of ridesharing in Denmark}.
\newblock \bibinfo{journal}{\emph{Transportation Research Part A: Policy and Practice}}  \bibinfo{volume}{78} (\bibinfo{year}{2015}), \bibinfo{pages}{113--123}.
\newblock
\showISSN{0965-8564}
\urldef\tempurl%
\url{https://doi.org/10.1016/j.tra.2015.04.033}
\showDOI{\tempurl}


\bibitem[Noguchi et~al\mbox{.}(2020)]%
        {noguchi2020personality}
\bibfield{author}{\bibinfo{person}{Yohei Noguchi}, \bibinfo{person}{Hiroko Kamide}, {and} \bibinfo{person}{Fumihide Tanaka}.} \bibinfo{year}{2020}\natexlab{}.
\newblock \showarticletitle{Personality traits for a social mediator robot encouraging elderly self-disclosure on loss experiences}.
\newblock \bibinfo{journal}{\emph{ACM Transactions on Human-Robot Interaction (THRI)}} \bibinfo{volume}{9}, \bibinfo{number}{3} (\bibinfo{year}{2020}), \bibinfo{pages}{1--24}.
\newblock


\bibitem[OpenAI(2023)]%
        {OpenAI_ChatGPT_4}
\bibfield{author}{\bibinfo{person}{OpenAI}.} \bibinfo{year}{2023}\natexlab{}.
\newblock \bibinfo{title}{ChatGPT 4.0}.
\newblock \bibinfo{howpublished}{\url{https://www.openai.com}}.
\newblock
\newblock
\shownote{Accessed: 2024-04-10}.


\bibitem[Paredes et~al\mbox{.}(2018)]%
        {paredes2018fast}
\bibfield{author}{\bibinfo{person}{Pablo~E Paredes}, \bibinfo{person}{Francisco Ordonez}, \bibinfo{person}{Wendy Ju}, {and} \bibinfo{person}{James~A Landay}.} \bibinfo{year}{2018}\natexlab{}.
\newblock \showarticletitle{Fast \& furious: detecting stress with a car steering wheel}. In \bibinfo{booktitle}{\emph{Proceedings of the 2018 CHI Conference on Human Factors in Computing Systems}}. \bibinfo{publisher}{ACM}, \bibinfo{address}{New York, NY, USA}, \bibinfo{pages}{1--12}.
\newblock


\bibitem[Pywell(1993)]%
        {pywell1993automotive}
\bibfield{author}{\bibinfo{person}{James~F Pywell}.} \bibinfo{year}{1993}\natexlab{}.
\newblock \bibinfo{booktitle}{\emph{Automotive seat design affecting comfort and safety}}.
\newblock \bibinfo{type}{{T}echnical {R}eport}. \bibinfo{institution}{Society of Automotive Engineers Technical Paper}.
\newblock


\bibitem[Rauh et~al\mbox{.}(2015)]%
        {rauh2015understanding}
\bibfield{author}{\bibinfo{person}{Nadine Rauh}, \bibinfo{person}{Thomas Franke}, {and} \bibinfo{person}{Josef~F Krems}.} \bibinfo{year}{2015}\natexlab{}.
\newblock \showarticletitle{Understanding the impact of electric vehicle driving experience on range anxiety}.
\newblock \bibinfo{journal}{\emph{Human factors}} \bibinfo{volume}{57}, \bibinfo{number}{1} (\bibinfo{year}{2015}), \bibinfo{pages}{177--187}.
\newblock


\bibitem[Schneider and Kiesler(2005)]%
        {schneider2005calling}
\bibfield{author}{\bibinfo{person}{Mike Schneider} {and} \bibinfo{person}{Sara Kiesler}.} \bibinfo{year}{2005}\natexlab{}.
\newblock \showarticletitle{Calling while driving: effects of providing remote traffic context}. In \bibinfo{booktitle}{\emph{Proceedings of the SIGCHI conference on human factors in computing systems}}. \bibinfo{pages}{561--569}.
\newblock


\bibitem[Seacrist et~al\mbox{.}(2018)]%
        {seacrist2018analysis}
\bibfield{author}{\bibinfo{person}{Thomas Seacrist}, \bibinfo{person}{Ethan~C Douglas}, \bibinfo{person}{Elaine Huang}, \bibinfo{person}{James Megariotis}, \bibinfo{person}{Abhiti Prabahar}, \bibinfo{person}{Abyaad Kashem}, \bibinfo{person}{Ayya Elzarka}, \bibinfo{person}{Leora Haber}, \bibinfo{person}{Taryn MacKinney}, {and} \bibinfo{person}{Helen Loeb}.} \bibinfo{year}{2018}\natexlab{}.
\newblock \showarticletitle{Analysis of near crashes among teen, young adult, and experienced adult drivers using the SHRP2 naturalistic driving study}.
\newblock \bibinfo{journal}{\emph{Traffic injury prevention}} \bibinfo{volume}{19}, \bibinfo{number}{sup1} (\bibinfo{year}{2018}), \bibinfo{pages}{S89--S96}.
\newblock


\bibitem[Semmens et~al\mbox{.}(2019)]%
        {semmens2019now}
\bibfield{author}{\bibinfo{person}{Rob Semmens}, \bibinfo{person}{Nikolas Martelaro}, \bibinfo{person}{Pushyami Kaveti}, \bibinfo{person}{Simon Stent}, {and} \bibinfo{person}{Wendy Ju}.} \bibinfo{year}{2019}\natexlab{}.
\newblock \showarticletitle{Is Now A Good Time?: An Empirical Study of Vehicle-Driver Communication Timing}. In \bibinfo{booktitle}{\emph{Proceedings of the 2019 CHI Conference on Human Factors in Computing Systems}}. ACM, \bibinfo{publisher}{ACM}, \bibinfo{address}{New York, NY, USA}, \bibinfo{pages}{637}.
\newblock


\bibitem[Seppelt et~al\mbox{.}(2018)]%
        {seppelt2018assessing}
\bibfield{author}{\bibinfo{person}{Bobbie Seppelt}, \bibinfo{person}{Sean Seaman}, \bibinfo{person}{Linda Angell}, \bibinfo{person}{Bruce Mehler}, \bibinfo{person}{Bryan Reimer}, \bibinfo{person}{T Victor}, \bibinfo{person}{M Bruyas}, \bibinfo{person}{M Regan}, \bibinfo{person}{C Brusque}, {and} \bibinfo{person}{A Fort}.} \bibinfo{year}{2018}\natexlab{}.
\newblock \showarticletitle{Assessing the effect of in-vehicle task interactions on attention management in safety-critical events}. In \bibinfo{booktitle}{\emph{6th Int. Conf. on Driver Distraction and Inattention}}. \bibinfo{pages}{1--11}.
\newblock


\bibitem[Sheller(2004)]%
        {sheller2004automotive}
\bibfield{author}{\bibinfo{person}{Mimi Sheller}.} \bibinfo{year}{2004}\natexlab{}.
\newblock \showarticletitle{Automotive emotions: Feeling the car}.
\newblock \bibinfo{journal}{\emph{Theory, culture \& society}} \bibinfo{volume}{21}, \bibinfo{number}{4-5} (\bibinfo{year}{2004}), \bibinfo{pages}{221--242}.
\newblock


\bibitem[Shen and V{\'e}rtiz(1997)]%
        {shen1997redefining}
\bibfield{author}{\bibinfo{person}{Wenqi Shen} {and} \bibinfo{person}{Alicia~M V{\'e}rtiz}.} \bibinfo{year}{1997}\natexlab{}.
\newblock \showarticletitle{Redefining seat comfort}.
\newblock \bibinfo{journal}{\emph{SAE transactions}} (\bibinfo{year}{1997}), \bibinfo{pages}{1066--1073}.
\newblock


\bibitem[Skjuve et~al\mbox{.}(2023)]%
        {skjuve2023chatgpt}
\bibfield{author}{\bibinfo{person}{Marita Skjuve}, \bibinfo{person}{Asbj\o{}rn F\o{}lstad}, {and} \bibinfo{person}{Petter~Bae Brandtzaeg}.} \bibinfo{year}{2023}\natexlab{}.
\newblock \showarticletitle{The User Experience of ChatGPT: Findings from a Questionnaire Study of Early Users}. In \bibinfo{booktitle}{\emph{Proceedings of the 5th International Conference on Conversational User Interfaces}} \emph{(\bibinfo{series}{CUI '23})}. \bibinfo{publisher}{Association for Computing Machinery}, \bibinfo{address}{New York, NY, USA}, Article \bibinfo{articleno}{2}, \bibinfo{numpages}{10}~pages.
\newblock
\showISBNx{9798400700149}
\urldef\tempurl%
\url{https://doi.org/10.1145/3571884.3597144}
\showDOI{\tempurl}


\bibitem[Smith et~al\mbox{.}(2010)]%
        {smith2010incivility}
\bibfield{author}{\bibinfo{person}{Philip Smith}, \bibinfo{person}{Timothy~L Phillips}, {and} \bibinfo{person}{Ryan~D King}.} \bibinfo{year}{2010}\natexlab{}.
\newblock \bibinfo{booktitle}{\emph{Incivility: The rude stranger in everyday life}}.
\newblock \bibinfo{publisher}{Cambridge University Press}.
\newblock


\bibitem[Stampf et~al\mbox{.}(2022)]%
        {10.1145/3546726}
\bibfield{author}{\bibinfo{person}{Annika Stampf}, \bibinfo{person}{Mark Colley}, {and} \bibinfo{person}{Enrico Rukzio}.} \bibinfo{year}{2022}\natexlab{}.
\newblock \showarticletitle{Towards Implicit Interaction in Highly Automated Vehicles - A Systematic Literature Review}.
\newblock \bibinfo{journal}{\emph{Proc. ACM Hum.-Comput. Interact.}} \bibinfo{volume}{6}, \bibinfo{number}{MHCI}, Article \bibinfo{articleno}{191} (\bibinfo{date}{sep} \bibinfo{year}{2022}), \bibinfo{numpages}{21}~pages.
\newblock
\urldef\tempurl%
\url{https://doi.org/10.1145/3546726}
\showDOI{\tempurl}


\bibitem[Suchman and Trigg(1995)]%
        {SUCHMAN1995233}
\bibfield{author}{\bibinfo{person}{Lucy~A. Suchman} {and} \bibinfo{person}{Randall~H. Trigg}.} \bibinfo{year}{1995}\natexlab{}.
\newblock \showarticletitle{Understanding Practice: Video as a Medium for Reflection and Design (Excerpt)}.
\newblock In \bibinfo{booktitle}{\emph{Readings in Human–Computer Interaction}}, \bibfield{editor}{\bibinfo{person}{RONALD~M. BAECKER}, \bibinfo{person}{JONATHAN GRUDIN}, \bibinfo{person}{WILLIAM~A.S. BUXTON}, {and} \bibinfo{person}{SAUL GREENBERG}} (Eds.). \bibinfo{publisher}{Morgan Kaufmann}, \bibinfo{pages}{233--240}.
\newblock
\showISBNx{978-0-08-051574-8}
\urldef\tempurl%
\url{https://doi.org/10.1016/B978-0-08-051574-8.50027-3}
\showDOI{\tempurl}


\bibitem[Tahir et~al\mbox{.}(2020)]%
        {tahir2020user}
\bibfield{author}{\bibinfo{person}{Yasir Tahir}, \bibinfo{person}{Justin Dauwels}, \bibinfo{person}{Daniel Thalmann}, {and} \bibinfo{person}{Nadia Magnenat~Thalmann}.} \bibinfo{year}{2020}\natexlab{}.
\newblock \showarticletitle{A user study of a humanoid robot as a social mediator for two-person conversations}.
\newblock \bibinfo{journal}{\emph{International Journal of Social Robotics}}  \bibinfo{volume}{12} (\bibinfo{year}{2020}), \bibinfo{pages}{1031--1044}.
\newblock


\bibitem[Tennent et~al\mbox{.}(2019)]%
        {tennent2019micbot}
\bibfield{author}{\bibinfo{person}{Hamish Tennent}, \bibinfo{person}{Solace Shen}, {and} \bibinfo{person}{Malte Jung}.} \bibinfo{year}{2019}\natexlab{}.
\newblock \showarticletitle{Micbot: A peripheral robotic object to shape conversational dynamics and team performance}. In \bibinfo{booktitle}{\emph{2019 14th ACM/IEEE International Conference on Human-Robot Interaction (HRI)}}. IEEE, \bibinfo{pages}{133--142}.
\newblock


\bibitem[Wada(2016)]%
        {wada2016motion}
\bibfield{author}{\bibinfo{person}{Takahiro Wada}.} \bibinfo{year}{2016}\natexlab{}.
\newblock \showarticletitle{Motion sickness in automated vehicles}. In \bibinfo{booktitle}{\emph{Advanced Vehicle Control: Proceedings of the 13th International Symposium on Advanced Vehicle Control (AVEC’16)}}. \bibinfo{pages}{169--174}.
\newblock


\bibitem[Wu et~al\mbox{.}(2021)]%
        {wu2021learningto}
\bibfield{author}{\bibinfo{person}{Tong Wu}, \bibinfo{person}{Nikolas Martelaro}, \bibinfo{person}{Simon Stent}, \bibinfo{person}{Jorge Ortiz}, {and} \bibinfo{person}{Wendy Ju}.} \bibinfo{year}{2021}\natexlab{}.
\newblock \showarticletitle{Learning When Agents Can Talk to Drivers Using the INAGT Dataset and Multisensor Fusion}.
\newblock \bibinfo{journal}{\emph{Proc. ACM Interact. Mob. Wearable Ubiquitous Technol.}} \bibinfo{volume}{5}, \bibinfo{number}{3}, Article \bibinfo{articleno}{133} (\bibinfo{date}{sep} \bibinfo{year}{2021}), \bibinfo{numpages}{28}~pages.
\newblock
\urldef\tempurl%
\url{https://doi.org/10.1145/3478125}
\showDOI{\tempurl}


\bibitem[Zhang et~al\mbox{.}(1996)]%
        {zhang1996identifying}
\bibfield{author}{\bibinfo{person}{Luian Zhang}, \bibinfo{person}{Martin~G Helander}, {and} \bibinfo{person}{Colin~G Drury}.} \bibinfo{year}{1996}\natexlab{}.
\newblock \showarticletitle{Identifying factors of comfort and discomfort in sitting}.
\newblock \bibinfo{journal}{\emph{Human factors}} \bibinfo{volume}{38}, \bibinfo{number}{3} (\bibinfo{year}{1996}), \bibinfo{pages}{377--389}.
\newblock


\bibitem[Zhang et~al\mbox{.}(2019)]%
        {zhang2019interaction}
\bibfield{author}{\bibinfo{person}{Ming Zhang}, \bibinfo{person}{Yuqi Zhang}, {and} \bibinfo{person}{Yazhuo Kong}.} \bibinfo{year}{2019}\natexlab{}.
\newblock \showarticletitle{Interaction between social pain and physical pain}.
\newblock \bibinfo{journal}{\emph{Brain Science Advances}} \bibinfo{volume}{5}, \bibinfo{number}{4} (\bibinfo{year}{2019}), \bibinfo{pages}{265--273}.
\newblock


\end{thebibliography}
